# The Internet of Battle Things

Alexander Kott, U.S. Army Research Laboratory

Ananthram Swami, Fellow IEEE, U.S. Army Research Laboratory

Bruce J. West, Fellow AAAS and APS, U.S. Army Research Office

The rapid emergence of Internet of Things is propelled by the logic of two irresistible technological arguments: machine intelligence and networked communications. Things are more useful and effective when they are smarter, and even more so when they can talk to each other. Exactly the same logic applies to things that populate the world of military battles. They too can serve the human warfighters better when they possess more intelligence and more ways to coordinate their actions among themselves. We call this the Internet of Battle Things, IoBT. In some ways, IoBT is already becoming a reality[1], but 20-30 years from now it is likely to become a dominant presence in warfare.

The battlefield of the future will be densely populated by a variety of entities ("things") – some intelligent and some only marginally so – performing a broad range of tasks: sensing, communicating, acting, and collaborating with each other and human warfighters[2]. They will include sensors, munitions, weapons, vehicles, robots, and human-wearable devices. Their capabilities will include selectively collecting and processing information, acting as agents to support sensemaking, undertaking coordinated defensive actions, and unleashing a variety of effects on the adversary. They will do all this collaboratively, continually communicating, coordinating, negotiating and jointly planning and executing their activities. In other words, they will be the Internet of Battle Things.

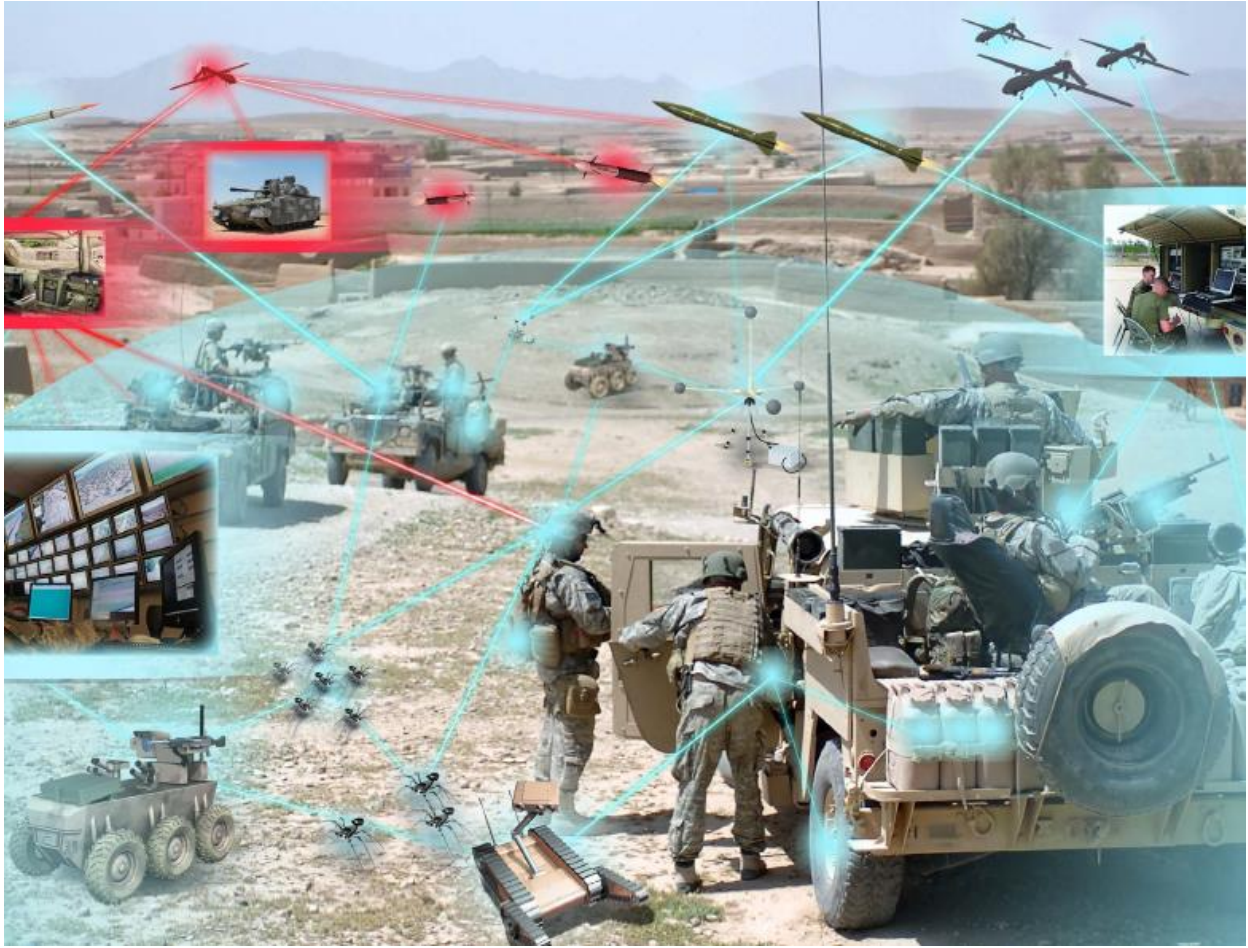

**Figure 1 Broad variety of systems and other "things" will communicate and collaborate on the battlefield. (Source: Illustration by Evan Jensen, U.S. Army Research Laboratory)**

To become a reality, however, this bold vision will have to overcome a number of major challenges. As one example of such a challenge, the communications among things will have to be flexible and adaptive to rapidly changing situations and military missions. This will involve organizing and managing large number of dynamic assets (devices and channels) to achieve changing objectives with multiple complex tradeoffs. Such adaptation, management and re-organization of the networks must be accomplished almost entirely autonomously, in order to avoid imposing additional burdens on the human warfighters, and without much reliance on support and maintenance services. How can this be done?

Secondly, human warfighters, under extreme cognitive and physical stress, will be strongly challenged by the massive complexity of the IoBT and of the information it will produce and carry. IoBT will have to assist the humans in making useful sense of this massive, complex, confusing, and potentially deceptive ocean of information, while taking into account the ever-changing mission, as well as, the social, cognitive and physical needs of humans.

Finally, nobody can discount the most important feature of the battle – the enemy. Besides being a lethal physical threat to the humans and IoBT, the enemy will be lurking in and around the IoBT networks and its information. IoBT itself will be a battlefield between its owners and defenders, and its uninvited part-owners – attackers. How will IoBT manage risk and uncertainty in this highly adversarial, deceptive environment?

These are some of the questions that were discussed at the strategic planning meeting that was organized by the US Army Research Laboratory (http://www.arl.army.mil) on 9-10 November 2015, and brought together a number of scientists from academia and industry, and military experts. The suggestions and concerns that emerged at the meeting coalesced into a rich and ambitious research agenda, summarized below.

## Managing and Adapting the IoBT

In spite of voluminous, current and past research on related topics in network science and engineering, merely by virtue of its exceptionally large scale IoBT will require new theoretical results, models, concepts and technical approaches. Indeed, IoBT's number of nodes for a future Army brigade might be several orders of magnitude greater than anything that has been considered in current practice. This is particularly true in the environments where such a brigade will find it advantageous to make use of networked devices and channels that it does not own, e.g., when making use of existing local civilian IoT (networking infrastructure and things) in military operations in a megacity. In this case, the meeting's participants suggested, IoBT scale on the order of a million of things per square kilometer is not an unreasonable target for exploration.

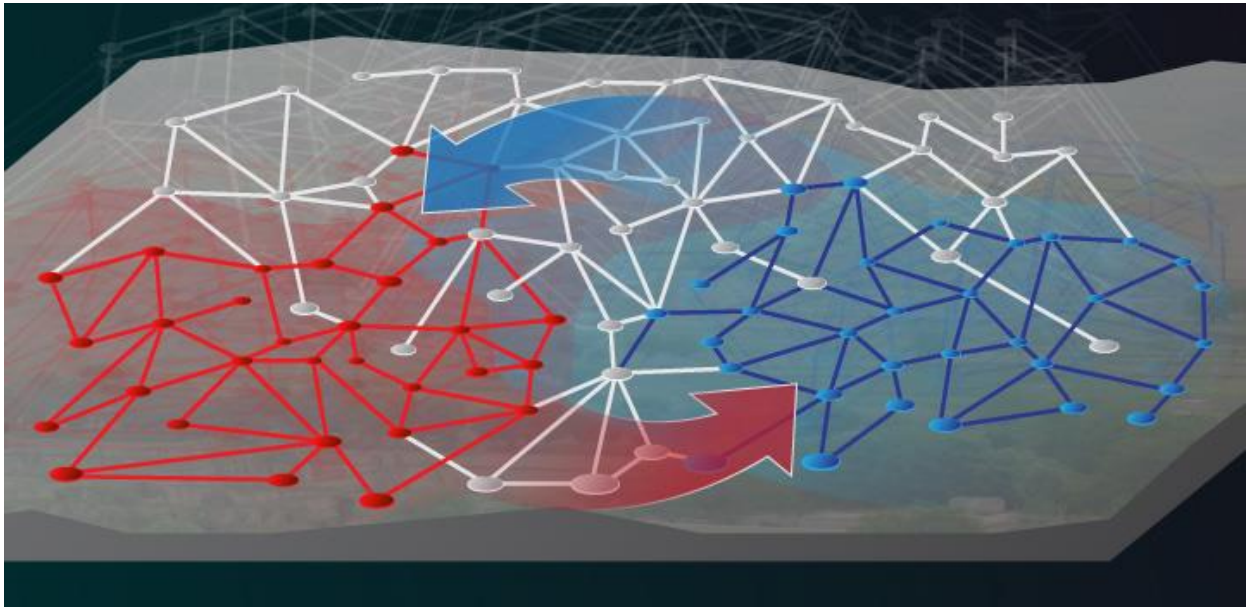

**Figure 2 Combatants will perform cyber attacks partly through the civilian Internet of Things to which they will be inevitably connected. (Source: Illustration by Evan Jensen, U.S. Army Research Laboratory)**

On the other hand, the massive scale of IoBT can be advantageous in practice and even for theoretical purposes. For example, availability of very large and densely positioned number of things, such as sensors can help eliminate currently common concerns about availability of any of them at a given time. To this end, theoretical results are needed to understand the degree of determinism resulting from very large ensemble of things and data.

Quite apart from its large scale, extreme heterogeneity of IoBT will call for new research and approaches. Not only the local IoT will consist of a broad range of commercial things and networks, but even the equipment that the warfighters will bring with them into the battle will likely rely on commercial offerings. It is probable that future commercial IoT will continue to exhibit a lack of standards, partly driven by desire of individual manufacturers to control its market, and will be generally chaotic. The military will have to adapt rapidly – and to have suitable technologies and techniques for such an adaptation – to use a broad variety of things, protocols, and communication technologies from multiple manufacturers.

In such a heterogeneous, highly dynamic and largely unpredictable environment, new approaches will be needed to facilitate discovery, characterization and tracking of relevant, available and useful things, dynamically in time and space. In particular, a military force that utilizes an existing IoT of a local society, e.g., a megacity, will not able to make reliable assumptions about behaviors and performance characteristics of any parts of its IoBT; instead such behaviors and characteristics will have to be learned and updated automatically and dynamically during the operation. Speaking of complex and unpredictable behaviors, one must not forget that humans – whether we call them "things" or not – are crucial and highly influential elements of IoBT. Behaviors and intents of humans – friendly warfighters, adversaries, and neutral civilians – will have to be dynamically detected, identified, characterized and projected in order to operate the IoBT.

Communications between things will also be challenged by high complexity, the dynamics and the scale of IoBT. Finding, sharing and managing communication channels, between large numbers of competing, heterogeneous and often unpredictable things will require novel approaches. Highly intelligent automation will be required to continually allocate and reconfigure the resources of the communications network. Information-sharing strategies and policies – who talks to whom, when, about what, and how long – will have to be automatically designed and modified dynamically. Highly scalable architectures and protocols will be needed, along with rigorous methods to determine and validate properties of protocols and architectures. In extreme situations, when IoBT experiences catastrophic collapse or becomes largely unavailable, or untrustworthy due to enemy actions, the autonomous management of IoBT will need to provide at least a "get me home" capability, which will enable the continuation of operations , even if at a limited level of functionality.

Additional complexity will arise from the wide range of timing constraints on communications. Some communications can wait for hours, while other communications will pose real-time requirements, for example for sensing and actuating. The channels will be constrained in highly heterogeneous ways as well. It is expected that 30 years from now, consumers will use wireless channels typically for only a few meters before the data enters fiber or other high-capacity channels; at the same time the military will require at least a few kilometers of wireless channels before encountering fiber.

To enable the dynamic management of IoBT, situational awareness of the IoBT as a whole will be formulated and updated rapidly and automatically; therefore new approaches will be desired directed towards the ability to measure relatively few variables of the complex system while thereby obtaining or inferring sufficiently complete information about the system.

While managing the IoBT, its purposes and uses must be taken into account, and these will be diverse. Some of its purposes will be relatively well understood, such as tactical military logistics or distributed computing. Others will be novel and will emerge from the availability of IoBT itself, such as perhaps use of IoBT for Position, Navigation and Timing (PNT) needs, and as a supplement to, or replacement for, GPS.

## Making IoBT Information Useful

As important as communications bandwidth is for effective operation of IoBT, it is the human cognition bandwidth that will emerge as the most severe constraint[3]. Human warfighters do not need and cannot process the enormously large flows of information produced and delivered by IoBT. Instead, humans seek well-formed, reasonably-sized, essential information that is highly relevant to their cognitive needs, such as effective indications and warnings that pertain to their current situation and mission. Responding to each thing that demands the human's attention, and to each piece of data that seem vaguely interesting, is not a feasible option in the context of IoBT. In fact, a key risk of IoBT is providing human warfighters (and intelligent things) with inappropriate information that leads – or misleads – to an action with an outcome worse than what would occur without that 'information'.

To make its information useful, IoBT technologies will have to deal with a large volume and complexity of information that are truly unprecedented in their extent[4]. Arguably, the quantity of data within IoBT will far exceed any likely advances predicted by Moore's Law and ever more efficient use of bandwidth might offer in the future. Besides the sheer volume, the complexity of the information will be formidable. For example, levels of abstraction, trustworthiness and value of information (produced or consumed) will vary drastically between different things.

The very foundations of information theory will need to be reconsidered; for example, ensemble probability densities are foundational for information theory, and require the underlying process to be ergodic. However, the IoBT is expected to have nonlinear dynamic processes that are sufficiently complex to generate events with non-ergodic statistics. The information entailed by the occurrence of such events must be based on single time series and not on an ensemble of time series[5]. Furthermore, non-intuitive, novel phenomena may emerge in the transfer of information between dissimilar large networks. An example would be in how situational awareness is modified by the information exchanged back and forth between IoBT and the social network of human warfighters, see e.g., [6] and references therein. Such unexpected phenomena may also influence – in yet-unknown ways – the ability of warfighters to control, inform and be informed by IoBT.

Still, at the very least, the IoBT's colossal volume of information must be reduced to a manageable level, and to a reasonably meaningful content, before it is delivered to humans and intelligent things. A likely target for compression and fusion of data into information, the meeting's participants conjectured, would be by a factor of 10E15. One approach to such a challenging fusion task is to populate IoBT, with a layered hierarchy of information brokers[7], or "concierges", which would aggregate, fuse, interpret and deliver appropriate information. The fusion process should begin at the lowest possible level; for example, whenever possible, all information-producing things should be equipped with the means to perform locally a degree of filtering, interpretation and fusion, before sending data to the network. Although such layers of intermediaries do complicate or restrict the discovery of underlying data, it may be a necessary price to pay for arriving at useful, manageable and meaningful information.

However, for information brokers to do their job, they need to know what constitutes useful information. Where would such knowledge come from? One source could be mission planning and rehearsal that could help determine what information is required by the mission-performing agents (human and artificial), and what is the likely available information. To capture the resulting knowledge, a machine-interpretable, formal, broadly applicable and military-relevant language will be needed for expressing information needs in highly heterogeneous IoBT[8]. Moving beyond the inevitable limitations of mission planning and rehearsal, IoBT will need approaches to self-learning of what information is needed for particular warfighter(s) and particular mission. Such approaches will likely require a form of integration of machine learning and semantic knowledge-based techniques.

More generally, executable models of the IoBT and its surrounding world are needed to enable validation, interpretation, fusion, and assessing trustworthiness of the information (e.g., [9]). Large scale simulation may help large scale sensing and interpretation of information in a targeted, purposeful manner. The research on formulating and automatically creating (and dynamically maintaining) such models is in its infancy. Effective solutions to this challenge will likely involve distributed self-modeling, self-calibration, and self-validation of IoBT.

## Dealing with Deception and Adversarial Nature of IoBT

Nothing differentiates IoBT from IoT more than the battle – the B in IoBT – against a determined and lethal enemy. The adversarial nature of the environment is the primary concern in the life of IoBT. The enemy threatens physical survival and functioning of IoBT by kinetic, directed-energy and electronic attacks against its things, by jamming the RF channels, by destroying fiber channels and by depriving IoBT of its power sources. The enemy also threatens the confidentiality, integrity, availability of the information within IoBT, by electronic eavesdropping, and by deploying malware into IoBT[10]. Finally, and perhaps most importantly, the enemy attacks the cognition of human warfighters. Humans will be elements within IoBT that are most susceptible to deceptions, particularly to those based on cognitive and cultural biases[11]. Humans' use of IoBT will be handicapped when they are concerned (even if incorrectly) that the information is untrustworthy[12] or that some elements of IoBT are controlled by the enemy. Similar susceptibilities, in part, apply to artificial intelligent entities.

Among the top priorities will be to minimize the enemy's opportunities to acquire information about IoBT and the warfighters it serves. While many of the applicable measures are the same as those for conventional battlefield networks, the exceptional scale, heterogeneity and density of IoBT offer additional opportunities for friendly information protection[13]. The sheer quantity of things (especially in those cases when friendly forces leverage the local IoT) permits the use of "disposable" security: devices that are believed to be potentially compromised by the enemy are simply discarded or disconnected from the IoBT. To defeat the enemy's eavesdropping, the defenders may want to take advantage of plentiful availability of things and inject misleading information into a fraction of them[14]. The density, complexity and diversity of message traffic within the IoBT will make it more difficult for the enemy to perform the traditional traffic analysis that could reveals details of the friendly command and control structure. Similarly, with a large number and density of things, it may be less expensive and more efficient to stymie the enemy's cyber intrusions by creating large, believable honeypots and honeynets, which are currently expensive to produce and to maintain dynamically. Although in the long run a honeynet may be less expensive than the devastation wrought by an adversary's cyber intrusion.

Besides acquiring friendly information, i.e., violating its confidentially, the enemy will attempt to violate the information's integrity, by modifying it with cyber malware, inserting rogue things into IoBT, intercepting and corrupting it while in motion, between the things, and presenting wrong information to the information-acquiring things, e.g., sensors. IoBT will likely fight back by anomaly detection that can highlight unexpected data patterns, unexplained dynamic changes, or lack of expected events (the dog that does not bark)[15]. To enable the anomaly detection, machine learning approaches will be developed to deal with the data as big and as dynamic as IoBT will possess. Such a continuous learning process will be computationally and bandwidth-wise expensive. It will be further challenged by the possibility that the enemy will adapt and evolve faster than the learning process can. In order to prevent the enemy from acquiring physical or software modifications of friendly things, approaches will be needed to achieve large-scale physical fingerprinting (e.g., collection of power consumption patterns) of things and continuous IoBT-wide monitoring of such patterns[16]. More generally, there will be means for active "stimulative intelligence" – ongoing physical and informational probing of IoBT that could help reveal the structure and behavior, including anomalous and suspicious ones, of the IoBT[17].

Learning normal patterns and detecting anomalous deviations, however, does not work well against a well-designed deception[18, 19]. In fact, learning can be a very dangerous double-edged sword with respect to deception. A common approach to deception is for the enemy to cause the friendly forces to learn a certain normal pattern, and then perform actions that blend into that pattern, but result in an unanticipated outcome. Any measure of normalcy can be defeated by effective deception. Still, very large scale and heterogeneity of IoBT may help defeat deception because "lying consistently is difficult;" it may be particularly difficult when the available sources of information are so numerous and are as heterogeneous as in IoBT. In general, much research is needed on approaches to counter-deception, discovery or rejection of deception for uniquely complex environment of IoBT[25]. And, considering that friendly IoBT will be necessarily connected with the local civilian IoT and thereby to the enemy's IoBT, approaches are needed to performing offensive operations executed within the intertwined space of friendly and enemy networks.

Such advanced capabilities will not be possible without new theoretical explorations. Fighting the battle of IoBT may require major new results in game theory, particularly focused on problems with very large number and very diverse game moves; near-infinite opportunities for probing; high complexity of utility functions, and partial observability of the game board limited to a very small fraction of the overall space. New theory is needed to formalize and normalize diverse definitions and conceptualizations of risks[20] and uncertainty. Deception should be integral to this theoretical analysis. For example, theoretical results should help predict the appropriate (or counterproductive) degree of complexity for a successful deception.

## Disclaimer
This article does not reflect the positions or views of the authors' employers.

## References

1. George I. Seffers, "Defense Department Awakens to Internet of Things," Signal Magazine, January 1, 2015

2. Kott, Alexander, David S. Alberts, and Cliff Wang. "Will Cybersecurity Dictate the Outcome of Future Wars?." *Computer* 48.12 (2015): 98-101.

3. Kranz, Matthias, Paul Holleis, and Albrecht Schmidt. "Embedded interaction: Interacting with the internet of things." *Internet Computing, IEEE* 14.2 (2010): 46-53.

4. Miorandi, Daniele, et al. "Internet of things: Vision, applications and research challenges." *Ad Hoc Networks* 10.7 (2012): 1497-1516.

5. B.J. West and P. Grigolini, *Complex Webs; Anticipating the Improbable*, Cambridge University Press, Cambridge, UK (2011).

6. B.J. West, M. Turalska and P. Grigolini, *Network of Echoes; Imitation, Innovation and Invisible Leaders*, Springer, NY (2014).

7. Korzun, Dmitry G., Sergey I. Balandin, and Andrei V. Gurtov. "Deployment of Smart Spaces in Internet of Things: Overview of the design challenges." *Internet of Things, Smart Spaces, and Next Generation Networking*. Springer Berlin Heidelberg, 2013. 48-59.

8. Barnaghi, Payam, et al. "Semantics for the Internet of Things: early progress and back to the future." *International Journal on Semantic Web and Information Systems (IJSWIS)* 8.1 (2012): 1-21.

9. Cho, Jin-Hee, Ananthram Swami, and Ing-Ray Chen. "A survey on trust management for mobile ad hoc networks." *Communications Surveys & Tutorials, IEEE* 13.4 (2011): 562-583.

10. Babar, Sachin, et al. "Proposed security model and threat taxonomy for the internet of things (IoT)." *Recent Trends in Network Security and Applications*. Springer Berlin Heidelberg, 2010. 420-429.

11. Kristin E. Heckman and Frank J. Stech, Cyber Denial, Deception and Counter Deception: A Framework for Supporting Active Cyber Defense. Springer 2015

12. Yan, Zheng, Peng Zhang, and Athanasios V. Vasilakos. "A survey on trust management for Internet of Things." *Journal of network and computer applications* 42 (2014): 120-134.

13. Satyajayant Misra, Reza Tourani and Nahid Ebrahimi Majd, "Secure content delivery in information-centric networks: design, implementation, and analyses," ACM SIGCOMM Information-Centric Networking Workshop, pp. 73–78, 2012.

14. Bisdikian, Chatschik, et al. "Trust and obfuscation principles for quality of information in emerging pervasive environments." *Pervasive Computing and Communications Workshops (PERCOM Workshops), 2012 IEEE International Conference on*. IEEE, 2012.

15. Raza, Shahid, Linus Wallgren, and Thiemo Voigt. "SVELTE: Real-time intrusion detection in the Internet of Things." *Ad hoc networks* 11.8 (2013): 2661-2674.

16. Roman, Rodrigo, Pablo Najera, and Javier Lopez. "Securing the internet of things." *Computer* 44.9 (2011): 51-58.

17. Ashraf, Qazi Mamoon, and Mohamed Hadi Habaebi. "Autonomic schemes for threat mitigation in Internet of Things." *Journal of Network and Computer Applications* 49 (2015): 112-127.

18. Vidalis, Stilianos, and Olga Angelopoulou. "Assessing identity theft in the Internet of Things." *IT CoNVergence PRActice (INPRA), 2 (1)* (2014).

19. Teixeira, André, et al. "A cyber security study of a SCADA energy management system: Stealthy deception attacks on the state estimator." *arXiv preprint arXiv:1011.1828* (2010).

20. Kott, Alexander, and Curtis Arnold. "The promises and challenges of continuous monitoring and risk scoring." *Security & Privacy, IEEE* 11.1 (2013): 90-93.